\documentclass[aps,10pt,twocolumn]{revtex4-1}
\usepackage[dvips]{graphicx}
\usepackage{subfigure}
\usepackage{bm}
\usepackage{color}
\usepackage{amsmath,amssymb}
\begin{document}
\title{
  Current-driven nucleation and propagation of antiferromagnetic skyrmionium
}
\author{S. A. Obadero$^{1,2}$}
\email{aobadero@aust.edu.ng}
\author{Y. Yamane,$^2$}
\author{C. A. Akosa$^{1,2}$}
\author{G. Tatara$^{2,3}$}
\affiliation{$^1$Department of Theoretical and Applied Physics, African University of Science and Technology (AUST), Km 10 Airport Road, Galadimawa, Abuja F.C.T, Nigeria}
\affiliation{$^2$RIKEN Center for Emergent Matter Science (CEMS), 2-1 Hirosawa, Wako, Saitama 351-0198, Japan}
\affiliation{$^3$RIKEN Cluster for Pioneering Research (CPR), 2-1 Hirosawa, Wako, Saitama, 351-0198 Japan}
\begin{abstract}
We present a theoretical study on nucleation and propagation of antiferromagnetic skyrmionium induced by spin current injection.
A skyrmionium, also known as $2\pi$ skyrmion, is a vortex-like magnetic structure characterized by a topological charge, the so-called skyrmion number, being zero.
We find that an antiferromagnetic skyrmionium can be generated via a local injection of spin current with toroidal distribution.
A spatially uniform spin current is then demonstrated to induce a propagation of the skyrmionium.
We derive an expression for the skyrmionium velocity based on a collective-coordinate model, which agrees well with our numerical results.
Our findings suggest a novel way of nucleating an antiferromagnetic skyrmionium, as well as an analytical estimation of the current-driven skyrmionium velocity. 
\end{abstract}
\maketitle
%
%
\section{Introduction}
Ever since the prediction\cite{Bogdanov,Rossler} and experimental discoveries\cite{Muhlbauer,Yu,Heinze,Seki,Raicevic} of magnetic skyrmions in magnetic materials with broken inversion symmetry, they have been generating increasing attention in the field of spintronics.
Skyrmions exhibit rich physics stemming from their characteristic, topologically nontrivial structures, and are deemed a promising player for future technological applications\cite{Fert,Nagaosa,Finocchio,Fert2017}.
Central to these applications is efficient manipulation of skyrmions in nanostructures;
several approaches with the use of electric current\cite{Jonietz,Yu2012}, spin waves\cite{Iwasaki-magnon,Schutte}, temperature gradient\cite{Kong,MochizukiT}, electric field gradient\cite{YHLiu}, and magnetic field gradient\cite{Komineas,SLZhang} have been demonstrated to control the dynamics of magnetic skyrmions. 
From a topological perspective, a skyrmion is characterized by a topological charge $Q=\pm1$\cite{Rajaraman}, known as skyrmion number.
A direct consequence of the nonzero $Q$ is that a moving magnetic skyrmion experiences a Magnus force\cite{Jiang,Litzius}, hampering a precise control of its motion.
There is thus a growing need to explore materials and magnetic structures where this undesirable effect can be overcome.

A skyrmionium (or 2$\pi$ skyrmion)\cite{Bogdanov1999,Finazzi,Komineas,Zhang,Shen,Gobel2019} refers to a magnetic texture that can be viewed as two nested skyrmions with opposite $Q$.
A skyrmionium thus carries $Q=0$ and is free from the Magnus force when driven into motion.
While both skyrmion and skyrmionium have been experimentally observed and investigated in ferromagnets, their antiferromagnetic counterparts have only been theoretically predicted recently\cite{Bogdanov2002,Barker,Hristo,Zhang-sk,Jin,Gobel,Collins,Fujita,Bhukta}.
The latter are also unaffected by the Magnus force, because it operates on the two different magnetic sublattices in the opposite sense, canceling out the effects of each other.
Antiferromagnetic materials have been widely projected as a viable replacement of ferromagnets deployed in some of the existing technologies due to their advantageous features\cite{MacDonald,Jungwirth,Baltz}; 
they are robust against external magnetic fields, produce no or negligibly small demagnetizing field, and display ultrafast dynamics.
Antiferromagnetic skyrmion/skyrmionium will therefore play pivotal roles in the next generation of spintronics applications.
Compared to the antiferromagnetic skyrmion, however, relatively little study has been devoted to the antiferromagnetic skyrmionium thus far\cite{Fujita,Bhukta}.

In this work, we propose a novel way of nucleating an antiferromagnetic skyrmionium, and investigate a current-driven propagation of skyrmionium. 
Micromagnetic simulation reveals that a local spin current injection with toroidal distribution can generate a skyrmionium.
We then derive the skyrmionium velocity driven by uniform spin current injection, based on a collective-coordinate model.
The numerically-observed skyrmionium motion agrees well with the analytical result, confirming the absence of Magnus force.

%
%
\section{Model}
We consider a two-dimensional antiferromaget (AFM) composed of two equivalent magnetic sublattices ($A$ and $B$) with constant saturation magnetization $M_{\rm S}$.
The spatial variation of the magnetic moments within each sublattice is assumed to be sufficiently slow compared to the atomistic length scale, so that the AFM can be appropriately modeled by performing a coarse graining for each sublattice\cite{Lifshitz}.
The classical vector field ${\vec m}_A ({\vec x}, t)$ $(|{\vec m}_A ({\vec x}, t)| = 1)$ represents the magnetization direction in the sublattice $A$, with a similar definition for ${\vec m}_B ({\vec x}, t )$.

In the continuous limit, we model the AFM by the following magnetic energy density $u$\cite{Bogdanov2002};
\begin{eqnarray}
  u  &=&   J_0 {\vec m}_A \cdot {\vec m}_B
             + A_1 \sum_{\zeta=A,B} \sum_{\mu=1,2}  \left( \frac{\partial{\vec m}_\zeta}{\partial x_\mu} \right)^2 \nonumber \\ &&
              - A_2 \sum_{\mu=1,2} 
                       \frac{\partial{\vec m}_A}{\partial x_\mu} \cdot \frac{\partial {\vec m}_B}{\partial x_\mu} 
              - K \sum_{\zeta=A,B} \left( {\vec m}_\zeta \cdot {\vec e}_3 \right)^2    \nonumber \\ &&
              - \frac{D}{4} \left( {\vec m}_A - {\vec m}_B \right) \cdot \left[ \left( {\vec e}_3 \times {\vec \nabla} \right) \times \left( {\vec m}_A - {\vec m}_B \right) \right]  ,
  \label{u}
\end{eqnarray}
where $ J_0 ( > 0 ) $ is the homogeneous exchange coupling energy, $A_1$ and $A_2$ are the isotropic exchange stiffnesses, $K(>0)$ is the uniaxial anisotropy constant along the $x_3$ axis, $D (>0)$ characterizes the Dzyaloshinskii-Moriya interaction (DMI), and ${\vec e}_\mu$ is the unit vector along the $x_\mu$ axis.

The dynamics of ${\vec m}_\zeta$ ($ \zeta =A, B $) are described by the coupled Landau-Lifshitz-Gilbert equations;
\begin{equation}
  \frac{\partial{\vec m}_\zeta}{\partial t}  =  - {\vec m}_\zeta \times \gamma {\vec H}_\zeta
                                                                  + \alpha {\vec m}_\zeta \times
                                                                                \frac{\partial{\vec m}_\zeta}{\partial t}  
                                                                 - {\vec m}_\zeta \times
                                                                                            \left( {\vec m}_\zeta \times {\vec p} \right)  ,                                                                               
   \label{llg}
\end{equation}
where $\gamma$ and $\alpha$ are the the gyromagnetic ratio and the Gilbert damping constant, respectively, (which are assumed for simplicity to be sublattice independent) and $ {\vec H}_\zeta =  - (\mu_0 M_{\rm S})^{-1} \delta u / \delta {\vec m}_\zeta$ is the effective magnetic field for the sublattice $\zeta$.
The last term in Eq.~(\ref{llg}) is the Slonczewski-Berger spin-transfer torque\cite{Slonczewski} due to spin current injection into the AFM, with ${\vec p}$ representing the magnitude and polarization of the spin current.
Here, we have assumed that the injected spin is transferred equiprobably to each of the sublattices.

In the present work, we consider the spin current injection realized via spin Hall effect in a nonmagnetic metal (NM)/AFM bilayer [Fig.~1.~(a)];
the spin current ${\vec j}_{\rm s}$ induced in NM due to spin Hall effect diffuses into the adjacent AFM and exerts the spin-transfer torque on the magnetizations\cite{Liu2011}.
(the spin-transfer torque of this mechanism is also called spin orbit torque or spin Hall torque.)
The directions of the electric current density ${\vec j}_{\rm c}$ in NM, the spin current flow ${\vec j}_{\rm s}$, and the polarization of the spin current ${\vec p}$ satisfy the relationship ${\vec j}_{\rm c} \propto {\vec j}_{\rm s} \times {\vec p} $.
The spatial distribution of ${\vec p} ({\vec x},t)$ is thus determined by that of ${\vec j}_{\rm c} ({\vec x},t)$, as ${\vec j}_{\rm s}$ is along the $x_3$ axis (the film-normal direction).
In the following sections, we will consider a spatially-nonuniform as well as uniform ${\vec p}$.
The magnitude of ${\vec p}$ carried by the injected spin current is given by 
\begin{equation}
  | {\vec p} | = \frac{ \gamma \hbar }{ 2 e \mu_0 M_{\rm S} } \frac{ \theta_{\rm SHE} }{ t_{\rm AFM} } | {\vec j}_{\rm c} |  ,
\end{equation}
where $\theta_{\rm SHE}$ is the spin Hall angle of NM, and $t_{\rm AFM}$ is the thickness of the AFM thin film (along the $x_3$ axis).

\begin{figure}
  \centering
  \includegraphics[width=8cm ,bb=0 0 642 768]{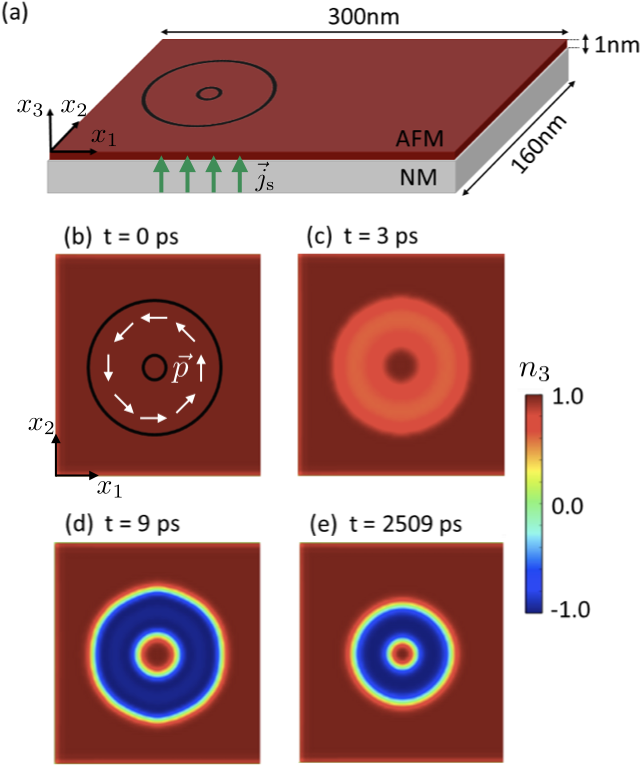}
  \caption{  (a) Schematic of our system;
                       bilayer of antiferromagnet (AFM) and nonmagnetic metal (NM). 
                       For the nucleation process, a 9 ps pulse of spin current ${\vec j}_{\rm s}$ (green arrows) is locally injected within the toroidal region indicated by the black circles, i.e., between the inner and outer circles, followed by a 2.5 ns of relaxation.
                  (b)-(e) Snapshots of the $n_3$ profile around the toroidal region ($0 \leq x_1 \leq 150$ nm and $0 \leq x_2 \leq 160$ nm) at selected times.
                      In (b), the spatial profile of ${\vec p}$ is schematically shown by white arrows.     
               }
  \label{fig01}
\end{figure}

We now introduce the N\'{e}el order parameter ${\vec n} ({\vec x}, t)$ and the ferromagnetic canting vector ${\vec m} ({\vec x}, t)$ by
\begin{eqnarray}
  {\vec n}  ({\vec x}, t)  &=&  \frac{{\vec m}_A  ({\vec x}, t) - {\vec m}_B  ({\vec x}, t) }{2} , \\
  {\vec m}  ({\vec x}, t)  &=&  \frac{{\vec m}_A  ({\vec x}, t) + {\vec m}_B  ({\vec x}, t) }{2}  ,
\end{eqnarray}
and the relations ${\vec n}^2 + {\vec m}^2 = 1$ and ${\vec n}\cdot {\vec m} = 0$ are a direct consequence of the above definitions.
Because the AFM exchange coupling is usually dominant over the other energies, one can safely assume $|{\vec m} ({\vec x}, t) |\ll 1$ and $|{\vec n} ({\vec x}, t) |\simeq1$.
By rewriting Eq.~(\ref{llg}) in terms of $({\vec n},{\vec m})$ and using $J_0 \gg D, K, |{\vec p}|$, one obtains, up to the first order of ${\vec m}$, the closed equation of motion for ${\vec n}$\cite{Helen2010}
\begin{equation}
  {\vec n} \times \left[ c^2 \Box {\vec n} 
                + 2 \gamma H_E \left( {\vec n} \times {\vec p} + {\vec \omega}_{\rm k,d} - \alpha \frac{\partial{\vec n}}{\partial t} 
                                                              \right)
                              \right]  = 0  ,
  \label{nlsm}
\end{equation}
while ${\vec m}$ is given by a function of ${\vec n}$ as
\begin{equation}
  {\vec m}  =  - \frac{1}{2\gamma H_E} {\vec n} \times \frac{\partial{\vec n}}{\partial t}  ,
  \label{m}
\end{equation}
where $H_E = J_0 / \mu_0 M_{\rm S}$, $c^2 = 2 \gamma^2 H_E ( 2A_1 + A_2 ) / \mu_0 M_{\rm S} $, $\Box = {\vec \nabla}^2 - (1/c^2) \partial^2 / \partial t^2 $, and ${\vec \omega}_{\rm k,d}$ is a quantity originating from the magnetic anisotropy and DMI\cite{Footnote}.
Equation~(\ref{m}) shows that, within our present approximation, ${\vec m}$ vanishes when $\partial {\vec n} / \partial t = 0$.

In equilibrium, therefore, the magnetic energy density in Eq.~(\ref{u}) is rewritten in terms only of ${\vec n}$ as
\begin{eqnarray}
  u &=&  
  \left( 2A_1 + A_2 \right) \sum_{\mu=1,2} \left( \frac{\partial{\vec n}}{\partial x_\mu} \right)^2  \nonumber \\ &&
  - D {\vec n} \cdot \left[ \left( {\vec e}_3 \times {\vec \nabla} \right) \times {\vec n} \right]
  - 2 K ( {\vec n} \cdot {\vec e}_3 )^2 .
  \label{u2}
\end{eqnarray}
This form of magnetic energy density is common to that of ferromagnets with exchange coupling, DMI and uniaxial anisotropy.
Equation~(\ref{u2}) thus indicates that, when the material parameters are appropriately chosen, a variety of magnetic textures that have been observed in the ferromagnets can also be formed by ${\vec n}$ in the AFM.
Antiferromagnetic skyrmionium is a vortex-like structure characterized by the skyrmion number $Q$ being zero;
\begin{equation}
  Q \equiv \frac{1}{4\pi} \int d^2x \ {\hat n} \cdot \left( \frac{ \partial {\hat n} }{ \partial x_1} \times \frac{ \partial {\hat n} }{ \partial x_2} \right) = 0 ,
\label{q}
\end{equation}
where $ {\hat n} = {\vec n} / | {\vec n} | $.

In the following, Eq.~(\ref{llg}) is numerically solved in terms of $({\vec m}_A, {\vec m}_B)$ to demonstrate nucleation and propagation of skyrmionium.
We also present an analytical investigation into skyrmionium propagation based on Eq.~(\ref{nlsm}).
For micromagnetic simulation, we consider an AFM rectangular thin film of dimensions $300\times160\times1$ nm$^3$ [Fig.~1~(a)], and divide it into $2\times2\times1$ nm$^3$ unit cells.
Both the two sublattice magnetizations ${\vec m}_A$ and ${\vec m}_B$ are defined at each unit cell in the spirit of the coarse graining.
As the model parameters, we employ $J_0 = 10^8$ J/m$^3$, $A_1=2.5\times10^{-12}$ J/m, $A_2=10^{-11}$ J/m, $D=4.2\times10^{-3}$ J/m$^2$, $K=4\times10^5$ J/m$^3$, $\alpha=0.2$, $M_{\rm S}=5.8\times10^5$ A/m, $\gamma=2.211\times10^5$ s$^{-1}$(A/m)$^{-1}$, and $\theta_{\rm SHE}=0.15$.

%
%
\section{Nucleation of Skyrmionium}
In Ref.~\onlinecite{Fujita}, a nucleation of antiferromagnetic skyrmioniums utilizing femtosecond optical vortex beams, via the heating effect, had been proposed.
We here study a skyrmionium nucleation induced by local application of an ultrafast current pulse, via the spin-transfer mechanism.
The configuration of our system for the nucleation is similar to that of Ref.~\onlinecite{Gobel2019}, except that they studied ferromagnetic skyrmioniums.

In the micromagnetic simulation, we start from the initial state with ${\vec n}$ homogeneous pointing in the $+x_3$ direction (except at the sample edges where the DMI causes the tilting of ${\vec n}$ from the $x_3$ axis).
The spin current $ {\vec j}_{\rm s} $ is locally injected within the toroidal region [indicated by two black circles in Fig.~1.~(a) and (b)] for 9 ps to excite the magnetization dynamics, followed by a 2.5 ns of relaxation without applying any external torque;
the center of the toroid is located at $x_1^0 = 75$ nm and $x_2^0 = 80$ nm, and the inner and outer radii of the toroid are 10 nm and 50 nm, respectively.
The spatial profile of ${\vec p}$ within the toroid is given by 
\begin{equation}
{\vec p} = | {\vec p} | ( - \sin\varphi {\vec e}_1 + \cos\varphi {\vec e}_2 )  ,
\end{equation} 
as indicated by white arrows in Fig.~1~(b), with $\varphi = \tan^{-1}\left[(x_2-x_2^0)/(x_1-x_1^0)\right]$, while ${\vec p} = 0$ outside the toroid.
Such a spin current injection corresponds to applying an electric current pulse in Corbino geometry, i.e., a radially symmetric  ${\vec j}_{\rm c} = j_{\rm c} ( \cos\varphi {\vec e}_1 + \sin\varphi {\vec e}_2 ) $ within the toroid region in NM.
Recently, it had been demonstrated that such an ultrashort electric pulse can be optically excited by utilizing a photosensitive switch\cite{Gobel2019,YYang}.
For the skyrmionium nucleation, we chose $ j_{\rm c} = 1.18\times10^{13}$ A/m$^2$.

In Fig.~1~(b)-(e), the snapshots of the $n_3$ profile around the center of the toroid at selected times are displayed.
At $t=3$ ps [Fig.~1~(c)], it is seen that the ``seed'' of a skyrmionium has been generated, where the in-plane components of ${\vec n}$ has been developed in the toroidal region.
The spin current injection is then completely turned off at 9 ps [Fig.~1~(d)], when ${\vec n}$ inside the toroidal region has flipped to the $-x_3$ direction.
After 2.5 ns of relaxation without current, the system reaches the state with a skyrmionium stabilized [Fig.~1~(e)].
As expected from the form of our DMI in Eq.~(\ref{u2}), the obtained skyrmionium has a N\'{e}el type structure\cite{Nagaosa}.

%
%
\section{Propagation of skyrmionium}
Now that we have successfully nucleated an skyrmionium, let us consider the propagation of it induced by a spatially uniform spin current injection (the inset of Fig.~2).
In Ref.~\onlinecite{Bhukta}, the skyrmionium velocity had been numerically estimated, finding its linear dependence on the current density.
Here, we address the skyrmionium propagation by an analytical as well as numerical approach.
Our analytical results are not only consistent with the earlier work, but also reveal the dependence of the skyrmionium velocity on the material and injected spin current.

First, we present a collective-coordinate model, where the rigid-motion approximation for the skyrmionium is employed, i.e., during its propagation the skyrmionium doesn't change its shape with respect to the coordinate frame co-moving with the skyrmionium.
The magnetization dynamics is thus fully described by the temporal evolution of the skyrmionium core position $(X_1, X_2)$;
${\vec n} ({\vec x}, t) = {\vec n} \left( {\vec x} - {\vec v} t \right) $, where ${\vec v} \equiv ( dX_1/dt, dX_2/dt ) $ is the skyrmionium velocity.
With this assumption, Eq.~(\ref{nlsm}) is reduced to a set of equations of motion for $(X_1,X_2)$ as
\begin{equation}
  M_\mu \frac{ d^2 X_\mu }{dt^2} + 2 \alpha M_\mu \gamma H_E \frac{ d X_\mu }{dt}  =  F_\mu , \label{eom} \\
 \end{equation}
 where $\mu=1,2$.
Here, the ``masses'' $M_\mu$ of the skyrmionium and the ``forces'' $F_\mu$ acting on it are defined by
\begin{eqnarray}
  M_\mu  &=&  \int d^2x \left( \frac{ \partial {\vec n} }{ \partial x_\mu} \right)^2  , \label{mass} \\
  F_\mu &=&  2 \gamma H_E \sum_j p_j I_{\mu j}   ,
\end{eqnarray}
where
\begin{equation}
  I_{\mu j}  =  \int d^2x \left( {\vec n} \times \frac{ \partial {\vec n} }{ \partial x_\mu } \right)_j  , \label{i}
\end{equation}
with $j = 1, 2, 3$.
Notice that in Eq.~(\ref{eom}) the dynamics of $X_1$ and $X_2$ are decoupled from each other, i.e., there appears no Magnus force or ``skyrmionium Hall effect.''

To compute Eqs.~(\ref{mass}) and (\ref{i}), we rely on the numerical profile of ${\vec n}$, since the analytical expression for a skyrmionium solution is unavailable.
With the equilibrium skyrmionium structure obtained in the previous section [Fig.~1~(e)], we find $M_1 \simeq M_2 \simeq 69.1 $ and $I_{12} \simeq - I_{21} \simeq - 4.4\times10^{-7} $ m;
we discarded in the spatial integral the region within 20 nm from the sample edges, not to take into account the magnetization tilting in this region, which has nothing to do with the skyrmionium motion.
For a N\'{e}el type skyrmionium, the other components of $I_{\mu j}$ are vanishingly small in general\cite{Gobel2019}.
A spin current with the polarization ${\vec p}$ along the $+x_1$ ($-x_2$) axis, therefore, drives a skyrmionium propagation in the $+x_2$ ($+x_1$) direction. [corresponding to $ {\vec j}_{\rm c} $ flowing in the $+x_2$ ($+x_1$) direction.]

Let us examine the skyrmioniun dynamics in the presence of spin current polarized in the $-x_2$ direction, i.e., ${\vec p} \parallel - {\vec e}_2$.
In this case, $F_2 = 0$ and thus $ dX_2 / dt = 0$, while a general solution $dX_1 / dt $ for Eq.~(\ref{eom}) is given by 
\begin{equation}
  \frac{dX_1 (t) }{dt}  =  v_1 \left( 1 - e^{ - 2 \alpha \gamma H_E t } \right)  ,
\end{equation} 
where the initial velocity being zero ($dX_1 / dt |_{t=0} = 0$) has been assumed, and $v_1$ is the terminal velocity defined by
\begin{equation}
  v_1 \equiv \left. \frac{dX_1}{dt} \right|_{t\rightarrow\infty}  =  \frac{ | I_{12} | }{\alpha M_1 } | {\vec p} |  .
\label{v1}
\end{equation} 
In Fig.~2, the skyrmionium velocities, $v_1$ [Eq.~(\ref{v1})] and $v_2(=0)$ in the $x_1$ and $x_2$ directions, respectively, are plotted by solid lines as functions of $j_{\rm c}$, where the same model parameters as before are used.


Now let us check the validity of the collective-coordinate model by comparing its predictions by micromagnetic calculations.
Starting from the relaxed skyrmionium [Fig.~1.~(e)], we simulate the skyrmionium propagation in the rectangular thin film with several different magnitude of ${\vec p} \parallel - {\vec e}_2$.
Numerically, the skyrmionium position $( X_1, X_2 )$ is measured by\cite{Komineas,Yamane2016}
\begin{equation}
  X_\mu  = \frac{ \int d^2x \ x_\mu \left( 1 - n_3 \right) f(n_3) }{ \int d^2x \left( 1 - n_3 \right) f(n_3) } , 
\end{equation}
where $f(n_3)=1$ when $| n_3 ({\vec x},t) | < 0.99 $ but otherwise zero.
The weighing function $f(n_3)$ is to ensure that only the magnetizations around the skyrmionium perimeter contributes to the averaging, otherwise leading to undesirable sample-size dependences of the evaluated quantities.
The numerical skyrmionium velocities are estimated from the displacement of $(X_1,X_2) $ during 280 ps (from $t=20$ ps to $t=300$ ps) under the spin current injection;
we discarded the first 20 ps not to take into account the transient magnetization dynamics right after the application of spin-transfer torque.

\begin{figure}
  \centering
  \includegraphics[width=8.5cm, bb=0 0 731 535]{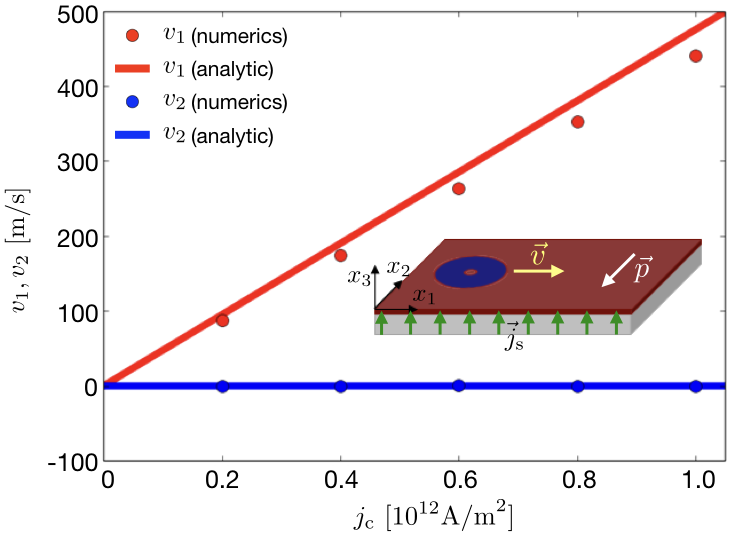}
  \caption{   Skyrmionium velocities $v_1$ and $v_2$, in the $x_1$ and $x_2$ directions, respectively, as functions of electric current density $j_{\rm c} $.
                  The analytical (numerical) results are represented by the solid lines (circles).
                  Schematically shown in the inset is the relation between the spin polarization ${\vec p}$ and the skyrmionium velocity ${\vec v}$.
                   For the skyrmionium propagation, the spin current injection is spatially uniform.
               }
  \label{fig02}
\end{figure}

In Fig.~2, the numerically estimated skyrmionium velocities are plotted by circles as functions of $j_{\rm c}$.
The micromagnetic simulation agrees semi-quantitatively with the collective coordinate model;
the skyrmionium velocity exhibits a linear dependence on the current density, and the relation between the directions of ${\vec p}$ and the skyrmionium propagation is consistent.
The quantitative discrepancy between the two approaches in $v_1$, which is less than 8 percent, may be attributed to the distortion of the skyrmionium shape.
At larger current density, the relative shift of the inner and outer skyrmions as well as the overall deformation becomes more appreciable, reducing the accuracy of the rigid-motion approximation.
We leave more systematic and complete investigations into the deformation effects to future work.

%
%
\section{Conclusions}
In conclusion, we have theoretically studied current-driven nucleation and propagation of antiferromagnetic skyrmionium.
Micromagnetic simulations have been performed to demonstrate that a local spin current injection with a toroidal distribution can nucleate a skyrmionium, and a spatially-uniform spin current can drive the skyrmionium into motion.
An analytical expression for the skyrmionium velocity derived based on a collective-coordinate model is consistent with the numerical result.
Our findings offer a novel way of nucleating an antiferromagnetic skyrmionium, as well as an estimation of the current-driven skyrmionium velocity for a given material.
%
%
\section{Acknowledgments}
This research was supported by Research Fellowship for Young Scientists (No. 17J03368) from Japan Society for the Promotion of Science (JSPS), Grant-in-Aid for Scientific Research (B) (No. 17H02929) from JSPS, and Grant-in-Aid for Scientific Research on Innovative Areas (No. 26103006) from the Ministry of Education, Culture, Sports, Science and Technology (MEXT), Japan.
S. A. O. acknowledges the sponsorship from the Pan African Materials Institute (PAMI).


\end{document}